\documentclass[twocolumn,groupedaddress,prb,floatfix,showpacs]{revtex4}
\usepackage{ulem}
\usepackage{graphicx} 
\usepackage{amsmath} 
\usepackage[english]{babel}  
\usepackage{graphicx}   
\usepackage{verbatim}
\usepackage{amsxtra}

\newcommand{\tx}{\tilde{x}}
\newcommand{\ty}{\tilde{y}}

\newcommand{\bn}{\begin{equation}}
\newcommand{\ee}{\end{equation}}
\newcommand{\bga}{\begin{eqnarray}}
\newcommand{\eda}{\end{eqnarray}}

\newcommand{\diff}{\mathrm{d}}
\newcommand{\eps}{\epsilon}

\newcommand{\A}{\text{\AA}}
\newcommand{\muf}{\mu_{\rm F}}

\newcommand{\chalmersMC}{$^{1}$ Department of Microtechnology and Nanoscience, MC2, Chalmers University of Technology,  SE-41296 G\"{o}teborg, Sweden} 

\begin{document}
\title{Polarization-balanced design of heterostructures: Application to AlN/GaN double-barrier structures}
\author{Kristian Berland}
\affiliation{\chalmersMC}
\author{Thorvald G. Andersson}
\affiliation{\chalmersMC}
\author{Per~Hyldgaard}
\affiliation{\chalmersMC}

\begin{abstract}
 Inversion- and 
 depletion-regions generally form at the interfaces between 
 doped leads (cladding layers) and the active region of polar heterostructures
like AlN/GaN and other nitride compounds.  
The band bending in the depletion region sets up a barrier which may seriously impede perpendicular electronic transport. This may ruin the performance of devices such as quantum-cascade lasers and resonant-tunneling diodes.  
Here we introduce the concepts of polarization balance and polarization-balanced designs:  A structure is polarization balanced when the applied bias match the voltage drop arising from spontaneous and piezeolectric fields.
Devices designed to operate at this bias have polarization-balanced designs. These concepts offer a systematic approach to avoid the formation of depletion regions.
As a test case, we consider the design of AlN/GaN double barrier structures with Al$_{\tilde{x}}$Ga$_{1-\tilde{x}}$N leads. To guide our efforts, we derive a simple relation between the intrinsic voltage drop arising from polar effects, average alloy composition of the active region, and the alloy concentration of the leads.
Polarization-balanced designs secure good filling of the ground state for unbiased structures, while for biased structures with efficient emptying of the active structure it removes the depletion barriers. 
\end{abstract}

\pacs{73.21.-b, 73.63.-b, 78.67.-n, 61.46.-w}

\maketitle

\section{Introduction}
Heterostructures of aluminium nitride (AlN) and gallium nitride (GaN) are key material systems for novel optoelectronic devices.\cite{FocusOnOpto,Machhadani:review}
The large band offset $\sim 2 {\rm eV}$ and direct-band gap
permits intersubband transitions in the near-infrared,\cite{intersub1,Intersubband:systematic,Theory:Transition,Absorption:superlattice,CoupledWells} which
makes them attractive for building devices working at fiberoptic wavelengths.
They hold the promise to extend the  operational regime of quantum cascade (QC) lasers both in the near and the far-infrared.\cite{designQCL,terahertz:Vukmirovi,terahertz:Wataru,Stattin:QCL} The latter due to the large LO-phonon energy $\omega_{LO} \sim 90 {\rm meV}$.
For wurtzite AlN/GaN structures grown in the c-plane direction, the band profile is distinguished by huge internal electrical fields. These fields arise from the interface charges originating in spontaneous and piezoelectric polarization.\cite{Heterojunction,QW:Macro}

\begin{figure}[t] 
  \begin{center}
    \includegraphics[width=8cm]{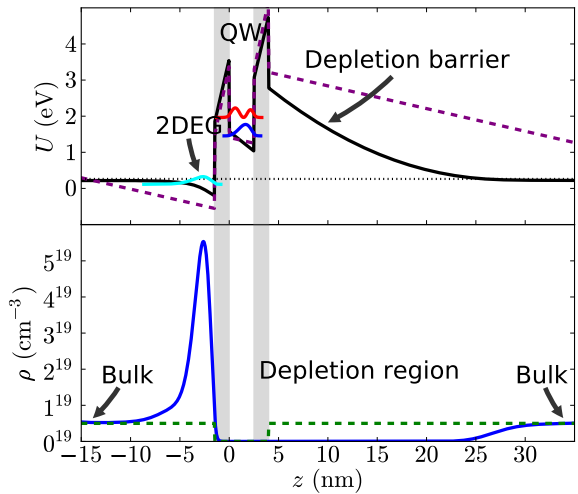}
  \end{center}
  \caption{Illustration of transport-inhibiting depletion barriers forming in typical heterostructures comprised of AlN and GaN layers. The figure shows conduction-band profile and bound states (upper panel), and charge density (lower panel). The active regions consist of 1.5 nm AlN barriers sandwiching a 2.5 nm GaN layer. Cladding layers consists of doped GaN $(5\cdot 10^{18} {\rm cm^{-3}})$.
The gray background indicates the structure. 
In the upper panel the full curves show the conduction-band profile and the dotted indicates the Fermi level. The wavy curves display the computed wave function squared of the three (quasi-)bound states of the structure. Their offsets are given their energy levels.  The dashed lines indicate a possible profile for an undoped heterostructure.  In the lower panel, the full curve shows the carrier density, while the dashed lines indicate the background doping density.}

  \label{fig:First}
  \end{figure} 
   
  AlN/GaN intersubband devices offer great opportunities, but challenges remain both for growth and design.
  Interface roughness,\cite{SeqTunn} high dislocation density, and scarcity of high-quality substrates limit device performance.
  Moreover, precise control of multiple quantum levels and efficient transport through the structure demand ultrathin AlN layers. This demand arise because the large band offset and heavy conduction electron mass $\sim (0.2-0.3) m_0$ weakens the coupling between states in adjacent quantum wells.
  The wide depletion regions and two-dimensional electron gases (2DEGs) forming at the two sides of an active region,\cite{Hermann2004,Intersubband:systematic,Machhadani:review,berland:tempads,IntersubbandOptics} exacerbates this challenge 
  for devices based on perpendicular quantum transport. For instance, in resonant-tunneling diodes and QC lasers, the depletion regions make it difficult to simultaneously obtain efficient current injection and control over the alignment of levels and transitions in the structure. 
  Presently, current-driven devices are mostly limited to resonant-tunneling diodes, often exhibiting large hysteresis.\cite{rtd:kikuchi,rtd1:Belyaev,rtd:pflugl,rtd:low_x,bayram:double_barrier,rtd:bayram2,SeqTunn}
A related issue exist for structures operating at zero bias (like absorption structures), especially those of few layers: as electrons accumulate in the 2DEG, the active region is depleted.\cite{Intersubband:systematic,berland:tempads,Machhadani:review,IntersubbandOptics}

To tackle some of the design problems, we here introduce 
the two related concepts of polarization balance and 
polarization-balanced design.  A structure is {\it polarization balanced} 
when the unscreened field vanish in the cladding layers (buffer and cap) layers. 
We will denote these regions as leads if the structure is intended for perpendicular transport. By unscreened we mean the field that arise
from the polarization-induced interface charges $\sigma_{\rm pol}$ and from the applied bias $V_{\rm bias}$ 
including dielectric screening by valence 
electrons, but excluding screening by conduction electrons and donors. The concept is similar in spirit 
to strain balance.\cite{strain:Harrison}  The electrical fields in the leads (cladding layers) generally 
bring about depletion and inversion regions outside the active region. Since the band tailing in the 
depletion region impedes electron transport, polarization balance 
helps enable efficient injection into the central active structure. Nonpolar structures are 
always polarization balanced at zero bias. Many polar structures are polarization balanced at a certain bias. 

{\it Polarization-balanced designs} are structures tailored to operate at a bias $V_{\rm bias}^0$ 
resulting in polarization balance. In this study, we rely on alloyed leads to retain a large design 
freedom for polarization-balanced designs. A few observations substantiated in this paper clarify the concept and highlight opportunities of such designs. First, these designs can, at a desired operational bias $V^0_{\rm bias}\ne 0$, provide flat-bands in the injection region and in the leads in general. There is no depletion barrier.  Second, polarization balance at $V^0_{\rm bias} = 0$ is also possible by using a suitable choice of composition in the cladding layers. In this case, charge accumulates within the active region. Third, if charge accumulates in the active region --- which can also happen for a finite bias, for instance (in a quasi-static picture), when the lowest level of the active region falls close to or below the Fermi level --- polarization balance does not result in flat bands. Fourth, a polar device can be operated for a bias before, at, or beyond polarization balance, that is 
$\varphi_{\rm bias}<\varphi_{\rm b}^0, \varphi_{\rm b}\approx \varphi_{\rm b}^0$, or $\varphi_{\rm b}>\varphi_{\rm b}^0$ with $\varphi_{\rm b}=-V_{\rm bias}$ and $\varphi^0_b=-V^0_{\rm bias}$. 
\footnote{At times we will use the symbol $\varphi_b=-V_{\rm bias}$ rather than the applied bias $V_{\rm bias}$ for convenience: using the standard convention for $z$-axis, we always consider an applied bias $V_{\rm bias}$ which is non-positive and a particle current going right-to-left.}
Fifth, nonpolar structures cannot operate before, and typically has to operate much beyond the bias resulting in polarization balance. 
Sixth, since in polar structures, current-driven devices such as quantum-cascade lasers can operate at the bias giving polarization balance, with ensuing flat bands in the injection region, good injection can be ensured without the need of tailoring graded digital alloys in the transport region. \cite{Faist1994}

Our calculations of conduction-band profile and 
quantum states presented here are based on the Schr\"odinger-Poisson (SP) equation\cite{Gunna2007,berland:tempads} in the 
effective-mass, envelope-function approximation, including non-parabolicity and exchange-correlation 
effects as described in Ref.~\onlinecite{berland:tempads}.

The plan of the paper is as follows. The second section motivates polarization-balanced designs. We illustrate the formation of depletion and inversion regions for an AlN/GaN heterostructure; we demonstrate that the depletion barrier seriously inhibit perpendicular current; and finally, we take a trial-and-error approach to polarization balance for AlN/GaN double-barrier structures with AlGaN leads of varying alloy concentrations.
In the third section, we develop theory to guide polarization-balanced designs. For an active region consisting of AlN and GaN layers strained to Al$_{\tx}$Ga$_{1-\tx}$N leads, a simple relation between intrinsic voltage drop, 
alloy concentration $\tx$, and average Al composition $x$ in the active region is developed.
The fourth section contains some simple applications: 
We first demonstrate that choosing a polarization-balance design for an unbiased structure secures good electron filling in the active regions; we then turn to a large parameter-space study of a two-level quantum cascade structure. We find that the use of AlGaN instead of GaN leads, significantly enlarge the freedom for realizing polarization-balanced designs. 
The final sections hold our discussions, prospects, and conclusions. 

\section{Depletion barriers in AlN/GaN heterostructures}

That depletion regions form outside the active regions in polar heterostructures is well known.\cite{Hermann2004,Leconte:vertical,Machhadani:review,berland:tempads,Berland:general} We here observe that these have a huge impact upon the function and transport properties of AlN/GaN heterostructures. By calculating the conduction-band profile of a double barrier we first illustrate the formation of such a depletion barrier. 
Then, by calculating the perpendicular current through a single AlN barrier having GaN leads, which allows for a simple and transparent analysis, we make evident how severely this barrier inhibits current injection. 
Finally, we bring to view how using AlGaN leads rather than GaN shortens or even removes these depletion regions. Such an approach opens the possibility of avoiding the depletion regions without needing huge biases that may prohibit realistic device design. This result motivates our theory developments and SP calculations in later sections. 

\subsection{Depletion and inversion regions}

Interface charges arise at the boundary between different layers of polar heterostructures as follows: 
\begin{equation}
  \sigma_{\rm pol}^i=\hat{z}\cdot(-\mathbf{P}^{i-1}+\mathbf{P}^i)=-P^{i-1}+P^i\,.
  \label{eq:polar_interface}
\end{equation}
The conduction band profile produced by these charges exhibit a characteristic sawtooth-like shape for a layered heterostructure.
\cite{gmachl:isb,Machhadani:review,berland:tempads,QW:Macro,Liu:Absorption,QW:Macro,Gunna2007}
If the lead composition and lattice constant is the same on both sides of the active region, so that $P_0=P_N$, the polar contributions to the electrical field vanish in the leads because the net charge produced by the polarization-induced charge in the active structure is zero: $\sum_i^N \sigma_{\rm pol}^i=0$.

The intrinsic voltage drop over the active structure, arising from the polarization-induced charges $V_{\rm pol}$, does not, in general, vanish. 
For a structure in equilibrium $V_{\rm bias}=0$ some mechanism must compensate for this potential drop, so that the Fermi level can remain constant throughout the structure.
For a doped structure, the electrical fields rearrange conduction electrons into an inversion (accumulation) region on one side of the structure, leaving behind a depletion (space-charge) region on the other. 
This charge transfer results in a potential drop that cancels the intrinsic voltage drop for a structure which is unbiased in equilibrium. If the screening by conduction electrons and ionized donors is so weak that the Fermi-level falls close to the valence band --- as for low donor concentration and large potential variations ---  valence electron may tunnel into the conduction band creating holes. Hence such polarization-induced doping can also compensate for the intrinsic potential drop.\cite{Induced2D} In our study, the formation of depletion and inversion regions in the conduction band is the sole mechanism responsible for securing equilibrium. 

Figure~\ref{fig:First} illustrates the formation of depletion and inversion regions for an unbiased double barrier structure, with AlN barriers, and GaN well and cladding layers. The conduction-band profile (upper panel) and electron density (lower panel) are obtained in a Schr\"odinger-Poisson calculation.\cite{berland:tempads} We use hard wall boundary conditions far outside the active region (boundaries not displayed) so that a proper bulk develops within the cladding layers. 
The lower panel shows that as electrons accumulate on the left, a wide depletion region forms ($\sim 25$ nm) on the right. The upper panel shows that the resulting band profile exhibits band bending in the inversion and depletion regions, and the characteristic sawtooth shape in the active part.
The ground state of the quantum well lies far above ($\sim 1{\rm eV}$) the Fermi level (dotted lines). In contrast, the energy level of the 2DEG falls below the Fermi level. In this inversion region, several quantised states exist, but electrons accumulate mostly in the 2DEG of lowest energy. \cite{2d:Ando,Gwo:ClassicalAndQuantum,Berland:general}
The dashed lines indicate the conduction-band profile for an undoped structure for some arbitrary  boundary condition. 
The central role played by the inversion and depletion regions for this simple heterostructure underlines the importance of considering the long-range screening as part of a device design.

\subsection{Tunneling barrier}

The wide depletion barriers seriously reduce tunneling current through a heterostructure. A qualitative WKB (Wentzel-Kramers-Brillouin) analysis serves as an illustration. The WKB tunneling probability is as follows:
\begin{equation}
  T_{\rm tun}(E) \approx e^{-\int_{\oslash}\diff z\, \sqrt{2m \left(U(z)-E\right)}/\hbar} = e^{-L_{\oslash} \langle {\sqrt{U-E}/ \hbar} \rangle_\oslash }.\label{eq:tunnprob}
\end{equation}
Here $\oslash$ identifies the classically forbidden regions and the angle brackets, an average over the forbidden region. $U$ is the potential energy, $E$, the energy, and $m$, the effective electron mass.  The tunneling current is given by the Tsu-Esaki formula:\cite{tsu_esaki}
\begin{align}
  I(V_{\rm bias})&=\frac{2m}{2\pi^2 \hbar^3} \int_0^\infty\diff E\, T^{V_{\rm bias}}_{\rm tun}(E)  D^{V_{\rm bias}}(E)\,,\label{eq:tsu_esaki} \\
D^V(E)& = \ln\left[\frac{1+\exp\left(\muf-E\right)/k_{\rm B}T}{1+\exp{(\muf-E-V)/k_{\rm B}T}}\right]\,.
\end{align}
If $V_{\rm bias}$ is large, a typical operational condition for AlN/GaN structures, electrons are not Pauli blocked (which enters through the support function $D^V(E)$ and Fermi level $\mu_F$). The tunneling current is then roughly proportional to the tunneling probability. 
We make two observations:
The heavy mass and large band offset of AlN/GaN reduce the tunneling probability compared to many other materials. 
And, a long classically forbidden region reduces the tunneling probability more than a high barrier does. In Eq.~\ref{eq:tunnprob} the term in the exponential increases linearly with the length, while the square root dampens the dependency on the barrier height. Thus, the barrier in the depletion region, which is typically of a wide spatial dimension, but of low energy, has a significant current-impeding effect. 

The current-voltage curve for a single AlN barrier with GaN leads is here calculated using a transmission coeffient obtained with a transfer-matrix method for arbitrary barrier shapes,\cite{ando:tunneling}  and the current $I$ with the Tsu-Esaki equation of Eq.~\ref{eq:tsu_esaki}.
This calculation further corroborates the WKB argument and attest the extreme current sensitivity to the depletion layer width. The AlN barrier is 1.0 nm wide (two monolayers). 
A positive background charge density of $N_d=5\cdot10^{18} {\rm cm^{-3}}$ accounts for doping of the GaN leads. The conduction-band profile $U(x)$ is obtained in an effective-mass Schr\"odinger-Poisson description.\cite{berland:tempads} A stepped Fermi-level induces the non-equilibrium charge-density distribution.

\begin{figure}[h] 
  \begin{center}
    \includegraphics[width=8cm]{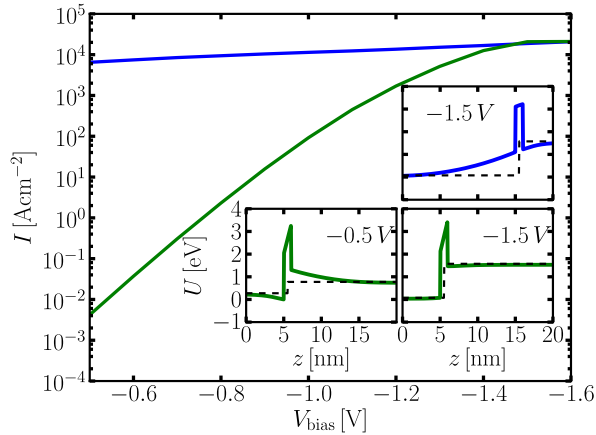}
  \end{center}
  \caption{Calculated current-voltage curve for a 1nm wide AlN barrier with GaN leads. 
 The lower green (upper) blue curve shows the current calculated with spontaneous and piezoelectric polarization included (neglected). The lower insets (upper inset) show the conduction band profile including (neglecting) these effects at a bias of $V_{\rm bias}=-0.5$ and $-1.5~{\rm V}\, (-1.5{\rm V})$. The dashed lines identifies the stepped effective Fermi-level.}
  \label{fig:singleBarrier}
\end{figure} 

Figure~\ref{fig:singleBarrier} displays the current-voltage curves for the barrier structure, calculated for conduction-band profiles obtained with and without the polar effects. The current increases enormously with bias until about $-1.4V$, for which the current in the polar case exceeds that of the nonpolar case (which does not exhibit this extreme sensitivity).  
The extreme bias sensitivity has also been documented in Refs~\onlinecite{Hermann2004} and ~\onlinecite{Leconte:vertical}. 
The two lower insets show the conduction-band profiles in the polar case for two sample voltages. It shows that for a voltage of $-0.5$V there is a wide depletion region, while for $-1.5$ voltage, the band is almost flat, with traces of an inversion region influencing the right side.
It is evident that the reduction in sensitivity coincides with achieving flat bands in the leads. 

The voltage drop arising from the polarization fields is $-1.38V$. Thus, for an applied bias of $V_{\rm bias}=-1.38 V$ the unscreened fields vanish, as the applied bias match the voltage drop over the barrier region. For the single barrier, the conduction band in the lead region, as obtained in the SP calculation, is approximately flat when the applied bias match the intrinsic voltage drop. 

Where the single barrier in a polar heterostructure represents an elementary case of polarization balance, a nonpolar heterostructure is the trivial case; Polarization-balance is achieved at zero bias. The upper inset in Fig. \ref{fig:singleBarrier} shows that, in the nonpolar case, the depletion and inversion regions that arise at nonzero bias do not inhibit current as the direction of the slope coincides with the movement of the electrons.\cite{spaceCharge,Brown}

\subsection{Biased double barrier structures}
\begin{figure*}[t]
  \begin{center}
    \includegraphics[width=16cm]{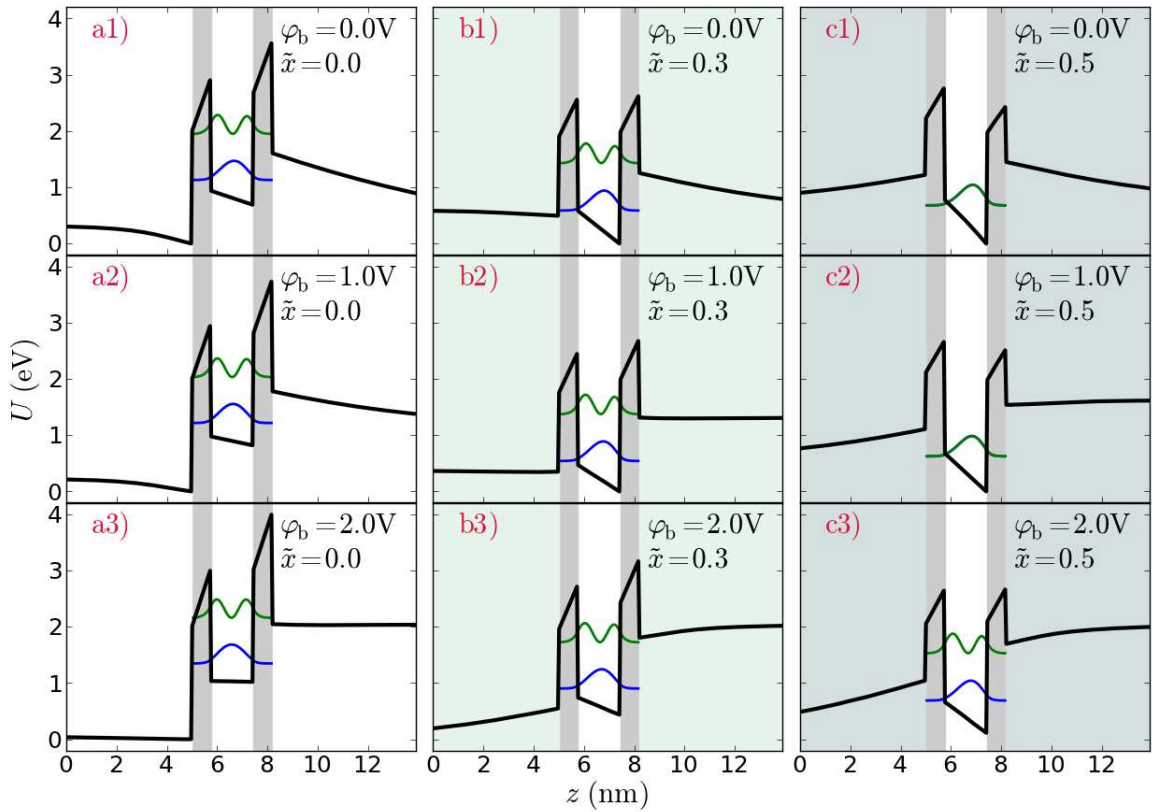}
  \end{center}
  \caption{Conduction-band profiles and bound states in quantum well of double barrier structures with 0.75 nm wide AlN barriers sandwiching an 1.7 nm wide GaN layer for varying bias $\varphi_{\rm b}=-V_{\rm bias}$ and Al concentrations $\tilde{x}$ in the doped  $(5\cdot10^{18} {\rm cm}^{-3})$ AlGaN leads.} 
\label{fig:3series}
\end{figure*}

Having studied a single barrier, we next study the conduction-band profile of double barrier structures which is arguable the simplest structures that exhibits an internal quantum structure within the active region. 
Its intended functionality is electroluminescence; the electron tunnel resonantly into the upper level of the quantum well, possibly make a transition to the lower level, and exit by tunnelling into the continuum. 
This toy structure allows us to exemplify design features of complex structures. The few parameters involved in its specification also permit brute-force trend studies, such as that presented in section 4. 

In designing many different structures, we found that a typical feature for AlN/GaN structures with GaN leads was that the bias required to overcome the depletion 
barriers, and thus get good transport conditions, was so large that it resulted in poor confinement of the upper level, and very hot electrons 
in the ejection lead. 
The left panels of Fig. \ref{fig:3series} illustrates these issues for a double-barrier structure. It displays the result of a Schr\"odinger-Poisson calculation for different applied biases, under the assumption of efficient emptying of the active region; the active region and the left lead have the same Fermi level. 
In the cases displayed in panels 'a1)' and 'a2)', a depletion barrier hinders electron injection. 
In panel 'a3)', the bias is large enough to give flat bands in the leads, and also within the quantum well. 
In this case, states in the injection lead resonate with the upper level of the active region. These energy levels match because the AlN layer is appropriately thin for a GaN well of this width. 
In polar heterostuctures, the energy levels depend strongly on the barrier widths, because of the typically large potential drops across the barrier.  
The 'a3)' structure exhibits troublesome features; the potential of the left barrier is so low that electrons easily escape prior to making a transition, and since the electron is confined in the triangular part of the barrier, increasing the well width does not help. That the electrons leaving the structure are hot can also be a problem. For periodic structures such as QC lasers, electrons may never relax before reaching the next period. 
It is also inefficient from a power-usage perspective. 

Using AlGaN leads with a finite Al concentration $\tilde{x}$, instead of GaN leads, reduces the potential drop over the active structure
at high-transport resonant-tunneling conditions since the unscrened field arise from the difference in polarization between the layers. 
As the quantum transition happens mostly within the active structure, the alloy scattering introduced should not severely broaden this peak. The middle and right panels, 'bx)' and 'cx)',  of Fig.~\ref{fig:3series} show band profiles calculated for alloyed leads. 
The 'b1)' panel shows that for zero bias and $\tilde{x}=0.3$, the depletion region is much reduced compared to $\tilde{x}=0.0$ in 'a1)'. 
By increasing the bias, as shown 'b2' panel, the band in the leads eventually becomes flat, requiring only about half the bias as with GaN leads as shown in 'a3)'. 
The 'b2)' structure confines the upper level much better than if the leads consist of GaN, and 
the electrons tunneling out of the structure remain fairly cold. Thus, it outperforms 'a3)'. 
Both 'b2)' and 'a3)' are polarization-balanced designs; they operate close to the flat-band condition. 

The 'c1)' structure highlights another important effect: Here, the energy of the lower level of the active region falls below the conduction band edge of the leads. This results in large band filling, with ensuing depletion regions on both sides of the barrier. Note that for the 'c2)' and 'c3)' structures, our assumptions about a stepped the Fermi level might be inappropriate. We include the structure for the purpose of illustration and completeness. 
We find that for this particular active structure, the use of Al$_{0.5}$Ga$_{0.5}$N leads [right panels / 'cx)'] prohibits flat band in the leads irregardless of applied bias.  

The conduction-band profiles of Fig.~\ref{fig:3series} demonstrate that we can remove the depletion barrier at, and above, a certain bias, and that this bias is lowered by using alloyed leads, unless electron accumulate within the active structure. Since flat band benefits electron injection, 
designing structures to operate at this bias is an excellent starting point for making current-driven devices where high power is essential. 
A strategy is therefore to restrict the design space to those structures which are polarization balanced. 
The next section develops theory for conditions that ensures an automatic restriction to such polarization-balanced designs. 

\section{Polarization balance}


\subsection{Intrinsic voltage drop}

The voltage drop over a heterostructure arising from polar effects is found by solving Poisson's equation. The solution is
\begin{equation} 
  V_{\rm pol} =  \sum_{i=1}^{N} (P_i-P_{\rm lead}) \frac{L_i}{\eps_i} \,. \label{eq:polbal1}
\end{equation}
Here $P_{\rm lead}$ is the polarization of the leads, assumed the same in both leads, while $P_i$ and $\eps_i$ is the polarization and dielectric constant of layer $i$. 
The polarization of a layer has both spontaneous and piezoelectric components: $P_i=P_i^{\rm SP}+P_i^{\rm piezeo}$. 

For systems with slowly varying composition or lattice constant, an integral formulation is more appropriate: 
\begin{equation}
  V_{\rm pol} =  \int_{\rm left\ lead}^{\rm right\ lead} \diff z\,\left[P(z)-P_{\rm lead}\right]\frac{1}{\eps(z)} \,. \label{eq:polbal2}
\end{equation} 
Naturally, this expression reduce to Eq.~\ref{eq:polbal1} for discrete variations of composition and lattice constant. 
Equations \ref{eq:polbal1} or \ref{eq:polbal2} determine the intrinsic voltage drop under broad conditions. Polarization balance is obtained when it equals the applied bias $V_{\rm pol}=V^0_{\rm bias}$. 
This condition generally requires that the two lead compositions are the same; otherwise, interface charges does not cancel out, and the unscreened fields cannot vanish in the lead region.  

If the active region is fully or partially relaxed, the lead regions on both sides should be defined so that they have the same lattice constant, and thereby piezoelectric contributions.

\subsection{Heterostructures of two-component alloys: AlN/AlGaN/GaN }
\begin{figure}[b] 
  \begin{center}
    \includegraphics[width=8cm]{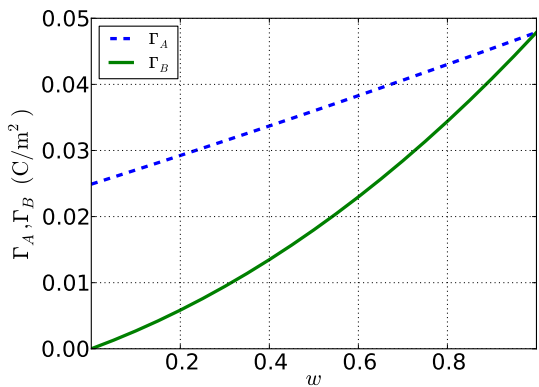}
  \end{center}
  \caption{Piezoelectric material parameters $\Gamma_A$ and $\Gamma_B$ as a function of the alloy concentration of an Al$_{w}$Ga$_{1-{w}}$N layer, 
as given by Eq.~\ref{eq:pz_par}.\label{fig_pz}}
\label{fig:DB_GaN}
\end{figure}

The expression for intrinsic voltage drop is here further developed for heterostructures with layers having alloy concentrations within the same class of binary alloys.
We use AlN/AlGaN/GaN layers as a specific case, but the expressions can be applied to other material systems. Here, Vegard's law determines the spontaneous polarization, $P^{\rm sp}(w)=P^{\rm sp}_{\rm GaN}(1-w) + P^{\rm sp}_{\rm AlN} w$, the bulk lattice constant, $a_{\rm bulk}(w)=(1-w)a_{\rm GaN} + w a_{\rm AlN}$, as well as other basic parameters of Al$_{w}$Ga$_{1-{w}}$N. The polarization balance concept is not restricted to this approximation, but computations are simpler and it allows for handy equations to be developed. 
Vegard's law is, for these quantities, a fair approximation for AlGaN layers, but it is less appropriate for alloys with larger bowing parameters, such as InGaN and InAlN.\cite{parameter_bowing}

The piezoelectric polarization can be expressed as
\begin{equation}
  P^{\rm piezo}_i (y_i,x_i)= y_i \Gamma^A(x_i) -\Gamma^B(x_i)\,,
  \label{eq:piezo}
\end{equation} 
where $x_i$ is the alloy concentration of the layer and $y_i$ is identified by the bulk alloy concentration corresponding to the lattice constant of the layer, that is
\begin{equation}
  a_{\rm bulk}(y_i)=a_i\,. \label{eq:bulk}
\end{equation}
The heterostructure may have a common lattice constant $x_i=x'$. The piezoelectric material parameters $\Gamma^A$ and $\Gamma^B$ are
\begin{align} 
\Gamma^A(w) &=  \left[a_{\rm AlN} - a_{\rm GaN} \right] \frac{2D(w)}{a(w)}\,,\nonumber \\
\Gamma^B(w) &=  \left[ a(w)-a_{\rm GaN}\right] \frac{2D(w)}{a(w)}\, \label{eq:pz_par}.
\end{align}
Here $ D(w)=e_{31}(w) - e_{33}(w) c_{13}(w)/c_{33}(w)$ with $c_{13}$ and $c_{33}$ being elastic coefficients and $e_{31}$ and $e_{33}$, piezoelectric coefficients.\cite{Gunna2007} In particular, we identify $\Gamma^A(0)\equiv \Gamma_{\rm GaN}$, $\Gamma^B(0)\equiv0$, and $\Gamma^A(1)=\Gamma^B(1)\equiv \Gamma_{\rm AlN}$. 
Figure~\ref{fig_pz} displays $\Gamma_A$ and $\Gamma_B$ as a function of alloy concentration $w$. While $\Gamma_A$ is almost linear, $\Gamma_B$ exhibits clear bowing. A linear approximation for this term is inaccurate. 

With the notation introduced here, we obtain
\begin{align}
V_{\rm pol} &=  \sum_{i=1}^{N-1} \left[ P^{\rm sp}_i + x_i \Gamma^A_i -\Gamma^B_i    - P_{\rm lead} \right]\frac{L_i}{\eps_i}\,, \nonumber  \\
P_{\rm lead} &=- \tx P^{\rm sp}_{\rm AlN}  - (1-\tx) P^{\rm sp}_{\rm GaN}+\ty \Gamma^A(\tx) -\Gamma^B(\tx)\,. \label{eq:Vint}
\end{align}
Here $\tx$ is the alloy concentration of the leads and $\ty$ is the bulk AlGaN concentration corresponding to the lattice constant of the leads (Eq.~\ref{eq:bulk}). 

Two consistency checks of Eq.~\ref{eq:Vint} are readily available: First, if  a layer (for instance the leads) is relaxed, $x_i=y_i$, its piezo-electric component, $P_{\rm piezo}(x_i,x_i)= x_i \Gamma^A(x_i) -\Gamma^B(x_i)=0$, vanish. Second, the intrinsic voltage drop does not depend on the starting point of the lead region: If we define part of a lead (layer $k$) as a layer within the active region, the intrinsic potential drop $V_{\rm pol}$ remains the same, because the additional $k$ term yields zero: $P_k-P_{\rm lead}=0$. 

\subsection{AlN/GaN layers strained to AlGaN leads}

If the active region is comprised of AlN and GaN layers, and these are strained to the leads ($x'=\tx=\ty)$, there is a simple relation between the 
lead alloy concentration $\tilde{x}$, intrinsic voltage drop over active region length $V_{\rm pol}/L$, and average concentration in the active region:
\begin{equation}
x=\sum L_i^{\rm AlN}/L. 
\end{equation}
The relation can be stated as follows: 
\begin{equation} 
\tx= \frac{-V_{\rm pol}/L + A x}{ A x  + B(1-x)} \,.
\label{eq:result}
\end{equation}
The material specific values $A$ and $B$ are given by
\begin{align}
  A&=\left( P_{\rm sp}^{\rm AlN}-P_{\rm sp}^{\rm GaN}-\Gamma^{\rm AlN} \right)\frac{1}{\eps_{\rm AlN}}\,,\\
  B&=\left(P_{\rm sp}^{\rm AlN}-P_{\rm sp}^{\rm GaN}-\Gamma^{\rm GaN}\right)\frac{1}{\eps_{\rm GaN}}\,, 
\end{align}
with materials-specific values listed in Table~\ref{tab:par}.
In deriving the relation (\ref{eq:result}) between intrinsic voltage drop and lead concentration,
we have used the Vergard's law as in the previous subsection. 

\begin{figure}[h] 
  \begin{center}
    \includegraphics[width=8cm]{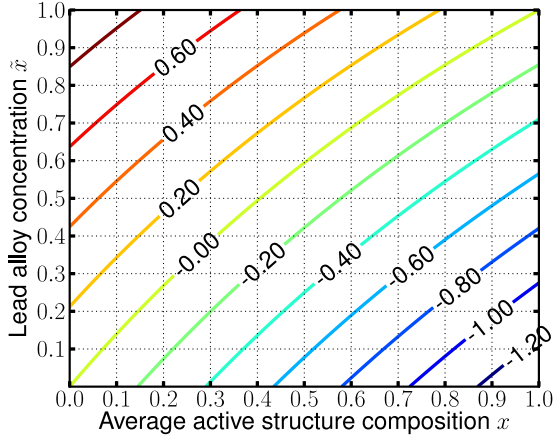}
  \end{center}
  \caption{The intrinsic voltage drop per length  $V_{\rm pol}/L$ [${\rm V}/{\rm nm}$] of the active region as a function of the average composition $x$
  of the active region and the alloy concentration of leads $\tilde{x}$.  The active region is strained to the leads. }
  \label{fig:cond}
\end{figure}

Figure ~\ref{fig:cond} displays the relation (\ref{eq:result}) between the average composition $x$ and 
lead allow concentration $\tilde{x}$ for different $V_{\rm pol}/L$. By setting the applied bias
according to these contours, we get polarization balance, that is, we remove the unscreened fields in the leads.  
 \begin{table}[h]
  \begin{ruledtabular}
  \caption{Polarization-related material quantities introduced here, based on the parameters of Ref.~\onlinecite{Vurgaftman2007}.}
\begin{tabular}{llllll}
$A~[\mathrm{V}/\text{\AA}]  $ & $B~[{\rm V/\A}]$ & $\Gamma_{\rm AlN}~[{\rm C/m^2}]$ & $\Gamma_{\rm GaN}~[{\rm C/m^2}]$ \\
-0.138 & -0.094 &  0.048   &  0.025
\end{tabular}

\label{tab:par}
\end{ruledtabular}
\end{table}

We confirm the consistency of Eq.~\ref{eq:result} by showing that two polarization balanced active regions are also naturally described as single active region. 
We consider two active regions with average compositions $x_1$ and $x_2$, of length $L_1$ and $L_2$ producing respectably a intrinsic voltage drop of $V_1$, $V_2$ for the same lead alloy concentration $\tx$. We test whether $\tx$ is the solution to the following equation (with unknown $z$) describing the two regions in terms of a single active region:
\begin{align} 
\nonumber &-\frac{V_1+V_2}{L_1+L_2} +A  \frac{x_1 L_1+x_2L_2}{L_1+L_2}  \\ =& z\left[ A \frac{x_1 L_1+x_2L_2}{L_1+L_2}   + B(1-  \frac{x_1 L_1+x_2L_2}{L_1+L2})\right]\,.
\label{eq:add}
\end{align}
Singling out the $L_1$ and $L_2$ terms, we obtain
\begin{align}
&L_1\left[-V_1/L_1 +Ax_1 -z\left(Ax_1+ B(1-x_1)\right)\right]+\nonumber\\ 
&L_2\left[-V_2/L_2 +Ax_2 -z\left(Ax_2+ B(1-x_2)\right)\right]=0\,.
\end{align}
Since $z=\tx$ is a solution, the two descriptions are equivalent.

\section{Examples of applications}

\subsection{Double barrier structure polarization balanced for zero bias}

\begin{figure}[h]
  \begin{center}
    \includegraphics[width=8cm]{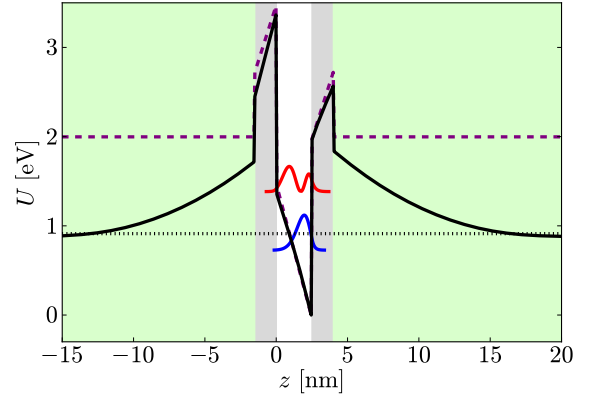}
  \end{center}
  \caption{Conduction-band profile and bound states of an optically active quantum well with a polarization-balanced design for zero bias. It consists of 2.5 nm wide AlN barriers, a 1.5 nm wide GaN well layer and cladding layers consisting of Al$_{0.63}$Ga$_{0.37}$N. Cladding layers are doped to $5\cdot 10^{18} {\rm cm^{-3}}$. The light gray [green] (white,gray) background indicate AlGaN (GaN,AlN). The dashed [purple] lines show the band profile of an undoped structure. The wavy curves indicate the wave function squared of the bound states offseted by their energy. The dotted line represents the Fermi level. }
  \label{fig:tails}
\end{figure}
Figure \ref{fig:tails} displays the conduction-band profile and bound states of an unbiased structure comprised of a GaN quantum well in-between AlN barriers and AlGaN cladding layers with alloy concentraction chosen for polarization balance  at zero bias. 
There is a significant band filling of the active region which generates depletion regions on both sides of the active region. This contrasts the depletion and inversion regions outside the depleted active region for GaN cladding layers, as displayed in Fig.~\ref{fig:First} and the '1a)' panel of Fig.~\ref{fig:3series}.
The dashed lines confirms the polarization balance equations, giving flat band in the cladding layers for an undoped structure.

The elimination of unscreened fields in the cladding layers (polarization balance) cause charging of the active region for unbiased structures because the energy of the lowest quantum well state falls below the condition band edge. 
The band bending in the cladding layers underline that merely fulfilling polarization balance does not necessarily result in a flat band. In the unbiased case polarization balance instead ensures efficient filling of the active region, rather than of the 2DEG at the interface. 
This result relates to an earlier observation that using the same alloy composition in the active region as in the cladding layers achieves good control over the electron population.\cite{Machhadani:review,IntersubbandOptics}

\begin{figure*}[t]
  \begin{center}
    \includegraphics[width=16cm]{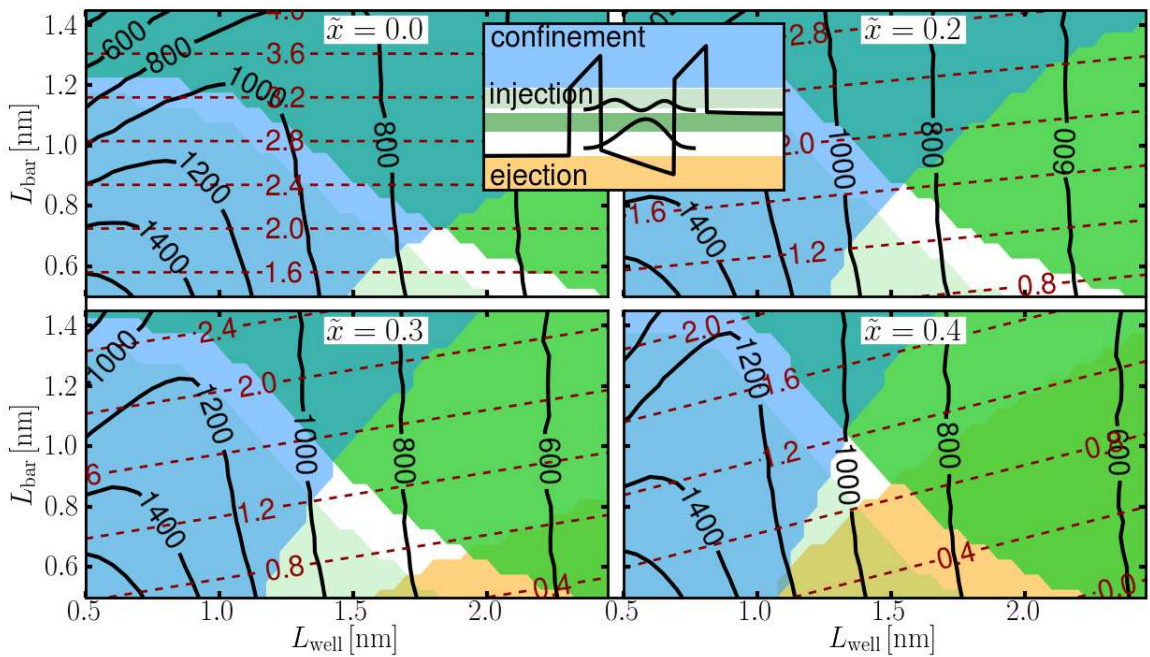}
  \end{center}
  \caption{(color) Design criteria, transition energy, and applied bias contours for polarization balanced two-level QC structure as a function of well and barrier widths for different Al$_{\tilde{x}}$Ga$_{1-\tilde{x}}$N alloy concentrations of the  leads: $\tilde{x}=0$, $0.2$, $0.3$, and $0.4$. 
  The numbers along the dashed curves mark the absolute value of applied bias ($V_{\rm bias}\,[V]$), set to achieve polarization-balance (using Eq.~\ref{eq:result}), while those on the full, mark the transition energy $\Delta E_{12}\,[{\rm meV}]$ between the two levels in the well.
    The white areas indicate where all design criteria are met; the shaded semi-transparent where they are not: The blue areas indicate where {\it confinement} of the upper level is poor. 
 The yellow, where {\it ejection} of the lower level is problematic. In the green (light green) areas, the {\it injection} is problematic because the energy of the upper level falls below (far above) the band edge of the injection lead.
The inset shows the band profile and the absolute-square of the localized wave-functions offset by their energy for the simple QC structure, and the shaded regions indicates design criteria.}
\label{fig:phases}
\end{figure*}

\subsection{Design of quantum-cascade structure}
If an active region is approximately charge neutral at operating conditions, the polarization-balanced design succeeds to remove depletion and inversion regions. This could be the case for heterostructures with efficient emptying of the active region. By adjusting the bias over a structure, we can always make a structure polarization balanced; however, the required bias can be unsuited for meaningful designs of the active structure. This was illustrated for GaN leads in our trial-and-error study detailed in Fig.~\ref{fig:3series}. 

The additional freedom of design gained by using alloyed leads for polarization balanced structures is here demonstrated with a large parameter study of 
a simple two-level quantum-cascade (QC) structure. The active region consists of two AlN barriers sandwiching a GaN well as in previously studied structures. We vary well $(L_{\rm w})$ and barrier $(L_{\rm bar})$ widths and lead alloy concentration $\tilde{x}$. The bias is always adjusted to obtain polarization balance. 
In calculating energies and wave functions, we only consider the active region. This is since we found only a tiny difference between the band profile as obtained 
with and without doping of the leads, given that 
charging of the active region is avoided.

The merits of a design depend on the intended application; for instance, gain and threshold current are key quantities for lasers. 
Rather than considering an application-specific assessment of our QC structure, we consider three intuitive, but somewhat crude, design criteria. The first is {\it injection} of the upper level: the energy should lie above the conduction band edge of the injection lead, but no more than 200 meV above. This upper bound is somewhat arbitrary; in practice it depends on the band filling of the leads and level broadening. However, the applied bias can be adjusted to accommodate slight misalignment. 
The second is {\it confinement} of the upper level,  the energy should fall below the triangular portion of the barrier to increase the fraction 
of the electrons making a transition to the lower level by avoiding the faster Fowler-Nordheim tunneling. The third is {\it  ejection} of the lower level: the energy should lie above the conduction band edge of the exit lead. Efficient ejection is essential to avoid charge-buildup causing depletion regions in the injection lead and Pauli blocking of optical transitions. 
Finally, we aim to design a structure operating at the fiberoptic frequency $\sim 800 {\rm meV}$.

Figure  \ref{fig:phases} details the results of our large parameter study of the simple QC structure. It shows that the structures with GaN leads does not fulfil the design criteria for a transition energy of 800 meV. The bias over the structure is also generally about twice the transition energy causing ultrahot electrons to exist in and around the structure. 
For larger alloy concentration in the leads, the acceptable parameter window moves to wider barriers and thinner wells. For $\tilde{x}=0.2$, and $\tilde{x}=0.3$, all design targets can be accommodated for the fiberoptic frequency. Increasing lead alloy concentrations further, the allowed window contracts, and for ${\tilde{x}=0.5}$ no parameters supports all design criteria.

\section{Discussion}

\subsection{Conduction-band regimes}

\begin{figure}[h] 
  \begin{center}
    \includegraphics[width=8cm]{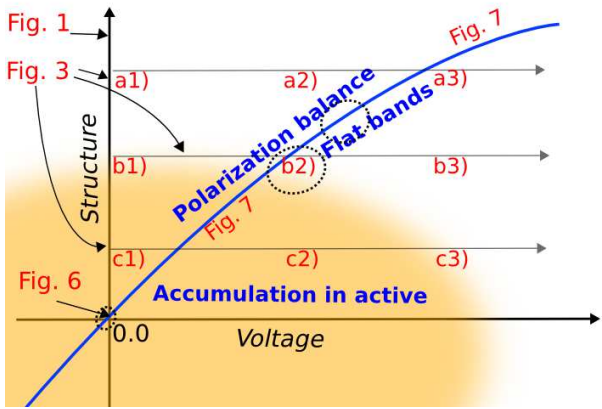}
  \end{center}
  \caption{Overview of considered structures: Different conduction-band regimes as function of applied voltage (horizontal-axis) and an abstract structure space (vertical-axis). The red curve marks polarization balance conditions. For structures along along this curve, the leads bands are flat unless charge accumulates within the active region, which happens where the curve enters the gray (orange) region. The figure number of structures considered in this study are marked on various spots. All cases considered in Fig.~7 are polarization balanced. Those within the dashed circles fulfil certain chosen design conditions. This makes them polarization-balanced designs. }
  \label{fig:overview}
\end{figure}

Figure~\ref{fig:overview} provides an overview of the different structures and bias conditions considered in this study and attempts to sort these into different conduction-band regimes. The horizontal-axis indicate different applied bias over the structure. The vertical-axis represents an abstract structure space; for instance, increasing the Al concentration of the cladding layer for a given active structure corresponds to moving downwards. The (blue) curve marks a structure in polarization balance. The gray (orange) region indicates that electrons accumulates within the active region, with ensuing depletion barriers on both sides of the active region.

We identify the position of structures studied here: 
In {\it Fig.~\ref{fig:First}}, we considered a structure far from polarization balance and with an empty active region. It exhibited a wide depletion region on the right side of the structure and a well-filled inversion region on the left. 
The panels of {\it Fig.~\ref{fig:3series}} displayed three different structures, under three different voltage conditions. The structures under varying bias is indicated by the three semi-transparent axes. Two of the panels ['a3'),'b2)']  showed structures under a bias that made them approximately polarization balanced. In two of the structures much charge accumulates within the active region ['c1)','c2)'] as the Fermi-level was above the lowest level of the active region.
The structure showed in {\it Fig.~\ref{fig:tails}} was designed for polarization balance at zero bias according to the theory developed in section 3. This secures excellent charge accumulation in the active region.
{\it Fig.~\ref{fig:phases}} displayed key results of numerous structures at a bias giving polarization balance. In {\it Fig.~\ref{fig:overview}} all these results scatter along the curve indicating polarization balance. Some of the designs resulted in filling of the active region (placing them within the gray region) and are therefore among the failed designs. The purpose of the dashed circles is to indicate (in an abstract sense) successful polarization-balanced designs according to the selected design criteria.

The placement of the results of the single barrier displayed in Fig.~\ref{fig:singleBarrier} are not shown here. We note that since the single barrier does not have any internal structure, accumulation within the active structure does not occur. A single barrier can not be designed for polarization balance at zero bias. 
The nonpolar case is always polarization-balanced at zero bias, but for such structures, polarization balance does not guarantee charge accumulation within the active region. 

\subsection{Prospects and challenges}

The theory developed for polarization balance may be useful for designing structures not intended as polarization-balanced designs, since it can also be used to identify the flat-band condition. 
Designing a device to operate at an applied bias larger than that resulting in polarization balance may be advantages for some purposes. For instance, to ensure an accumulation layers in the injection lead, making its operation more like that of a nonpolar structure.\cite{Bistability} 
Nevertheless, if an effective continuum should arise in the injection 
region, the bias across it should be small, to avoid the formation of a Stark-ladder.\cite{Wacker,Esaki} 

The huge strain in the structures considered in this study might render them unsuitable for actual applications, although there are numerous of attempts of making resonant-tunneling diodes using GaN leads which have even larger strains.\cite{rtd:kikuchi,rtd1:Belyaev,rtd:pflugl,bayram:double_barrier,rtd:bayram2} Using AlGaInN layers could be helpful, but complicates growth. With this additonal freedom, one could design structures that are both polarization and strain balanced.\cite{strain:Harrison} Only in special accidental cases can one fulfil both these criteria using layers within the same class of binary alloys (such as AlGaN). 

Lattice matching by using AlInN barriers and GaInN wells\cite{InAlGaN} removes strain completely. However, with only two different alloy compositions of the layers the generic polarization balance approach is impossible, since it in general requires layers of at least three different compositions. 
We therefore mention that polarization balance concepts can also be developed for leads that are superlattices. Such an approach is in the spirit of digital alloys. It would allow for structures that are both polarization balanced and lattice-matched (or the more relaxed strain-balance condition), for instance by combining Al$_{82}$In$_{18}$N and GaN.\cite{lattMatch1,polEngineering} Since strain is zero, we only balance the spontaneous part of the polarization.

Aside from presenting a solution to the current inhibiting effect of band tailing in the depletion region, polarization-balanced designs present an additional advantage: The shape of the conduction band does not depend on inversion and depletion regions. This has the consequence that 
the conduction-band profile essentially decouples from the specific doping and carrier concentration.
This reduces the importance of an exact control over the donor density for the functionality of a device, which is also helpful since the number of ionized donors is complicated to calculate in heterostructures, or when the binding energy is similar to the thermal energy.\cite{Theory:band_tails}
We note that in our quasi-static study, effects like charge trapping and formation of field domains\cite{Wacker} are not included, and these could complicate design considerations. 

For emission structures of short optical wavelength, and in particular for QC lasers made up of an alternating series of
optically active and interjecting current injections regions, the possibility to directly engineer zero-bias conditions in the injection region (or in the leads)  points to an additional strong advantage of polar heterostructures with polarization-balanced designs. In
contrast to conventional nonpolar structures, it is not necessary to grade the injection layers to compensate for the field.
A large applied bias does not inevitably result in a large electrical fields in the leads or, for that sake, in all of the active (transport-limiting) region. In a good design, the potential drop can be engineered to occur primarily in the region where the optical transition takes place.  
We conjecture that this advantage can result in a significant simplification of
the tailoring of the quantum wells and barriers that ensures efficient injection in devices like QC lasers.

\section{Conclusion}
The polarization balance concept represents a guiding principle for robust designs of wurtzite AlN/GaN heterostructures grown in the c-direction.
For high-power current-driven devices, it helps eliminate the depletion barrier impeding the perpendicular current. It also simplifies design since one can, to a larger extent, focus on the active structure. This is because the influence of the doping levels on the conduction-band profile is substantially reduced as the conduction band is flat in the leads. Furthermore, the flat bands in the leads makes grading of injection regions in QC lasers unnecessary. 
For such designs of unbiased structures it instead secures a large electron population in the active region. 

Our study has been restricted to AlGaN, since the large mass and band offset accentuates the current inhibiting effect of the depletion regions. However, the concept naturally generalize, not only to nitrides\cite{polEngineering} including InN and BN and their alloys, but also to heterostructures of other polar materials such as ZnMgO/ZnO.\cite{ZnMgO}

\section{Acknowledgement}
We thank F. Capasso and T. Ive for useful discussions. SNIC is acknowledged
for supporting KB's participation in the National Graduate School in Scientific Computing (NGSSC). Vinnova ({\it Banbrytande IKT}) provided financial support .

\bibliographystyle{apsrev}

\end{document}